\documentclass[12pt]{article}
\usepackage{graphicx}
\usepackage{epsfig}
\usepackage{amssymb,amsmath}
\usepackage{pstricks}
\usepackage{epstopdf}

\input{epsf.sty}

\newcommand{\be}{\begin{equation}}
\newcommand{\ee}{\end{equation}}
\newcommand{\ba}{\begin{eqnarray}}
\newcommand{\ea}{\end{eqnarray}}
\newcommand{\bi}{\begin{itemize}}
\newcommand{\ei}{\end{itemize}}

\newcommand{\nn}{\nonumber}

\renewcommand{\vec}[1]{{\bf #1}}

\newcommand{\RR}{{\rm I\kern -.2em  R}} 
\newcommand{\eq}{Eq.~}

\newcommand{\fig}{Fig.~}

\def\lsi{\raise0.3ex\hbox{$<$\kern-0.75em\raise-1.1ex\hbox{$\sim$}}}
\def\gsi{\raise0.3ex\hbox{$>$\kern-0.75em\raise-1.1ex\hbox{$\sim$}}}
\newcommand{\lsim}{\mathop{\lsi}}

\begin{document}

\begin{titlepage}

\begin{flushright}
MS-TP-09-25
\end{flushright}
\begin{centering}
\vfill
 

{\bf\Large The deconfinement transition of finite density QCD with heavy quarks from 
strong coupling series}

\vspace{0.8cm}
 
Jens Langelage and Owe Philipsen

\vspace{0.3cm}
{\em 
Institut f\"ur Theoretische Physik, Westf\"alische Wilhelms-Universit\"at M\"unster, \\
48149 M\"unster, Germany}

\vspace*{0.7cm}
 
\begin{abstract}
Starting from Wilson's action, we calculate strong coupling series for the 
Polyakov loop susceptibility in 
lattice gauge theories for various small $N_\tau$ 
in the thermodynamic limit. Analysing the series with Pad\'e approximants,
we estimate critical couplings and exponents for the deconfinement phase transition. 
For $SU(2)$ pure gauge theory our results agree with those from Monte-Carlo simulations 
within errors, which for the coarser $N_\tau=1,2$ lattices are at the percent level.
For QCD we include dynamical fermions via a hopping parameter expansion.
On a $N_{\tau}=1$ lattice with $N_f=1,2,3$, we
locate the second order critical point where the deconfinement transition turns into
a crossover. We furthermore  
determine the behaviour of the critical parameters with finite chemical potential
and find the first order region to shrink with growing $\mu$.
Our series moreover correctly reflects 
the known $Z(N)$ transition at imaginary chemical potential.
\end{abstract}
\end{centering}
 
\noindent
\vfill
\noindent
 

\end{titlepage} 

\section{Introduction}

Lattice Monte Carlo studies of the QCD 
phase diagram at finite temperature are still difficult for 
realistic quark masses, and at finite baryon density
are beset by the sign-problem. 
Despite the progress made in the last few years in circumventing these 
obstacles, such calculations still suffer severe restrictions. 
In particular, presently available methods are reliable only for
small quark chemical potentials, $\mu\lsim T$ \cite{op}. 
These difficulties motivate the search for 
alternative ways to learn about the phase diagram. A popular choice is to consider
lattice QCD in the strong coupling limit and to study its phase diagram
either by analytic mean field methods \cite{Kawamoto} or indeed by Monte Carlo evaluation
of the strongly coupled theory \cite{km},\cite{dff}, since the sign problem in this case
is much milder.
In the early days of lattice gauge theory, strong coupling expansions for zero 
temperature Yang-Mills theory have led to some analytical insights into field 
theories on a lattice (\cite{Wi},\cite{Mu} and \cite{drouffe} for a review). 
They have also been used to investigate finite temperature effects 
\cite{PS}-\cite{Su} 
with reasonable qualitative predictions,
but mostly using the crude approximation of neglecting spatial plaquettes altogether, 
cf.~\cite{Sv} and references therein for a review of early investigations. 
However, the strong coupling limit is far from the physical continuum 
theory and the lessons learned in this way are qualitative at best. There have also been 
attempts to go beyond the strong coupling limit \cite{Gr}-\cite{nakano}, mostly 
in mean field theory.
 
In this paper we address the question whether it is possible to make predictions
for the location and nature of the deconfinement
phase transition more quantitative by taking corrections into account,
and we can answer in the affirmative.
In previous work \cite{La} we have already shown for $SU(2)$ Yang-Mills theory 
that it is quite possible 
to include spatial plaquettes and to obtain series of several orders. 
For lattices with temporal extent $N_\tau=1-4$, 
this lead to satisfactory quantitative results for 
the equation of state up to the phase transition region. 
However, while the 
determination of the critical parameters was consistent with Monte Carlo 
results and universality arguments, 
it remained rather imprecise quantitatively. 
Here we considerably improve on this by using the Polyakov loop susceptibility as 
an observable, rather than the free energy. 
At a second order phase transition in infinite volume, 
this observable develops a singularity which is
well modelled by Pad\'e approximants to its series expansion, thus allowing to extract
the critical coupling and exponent in satisfactory agreement with numerical results
from \cite{BA},\cite{Fi},\cite{Ve}.

\begin{figure}
\vspace*{-1cm}
\begin{center}
\scalebox{0.4}{
\includegraphics{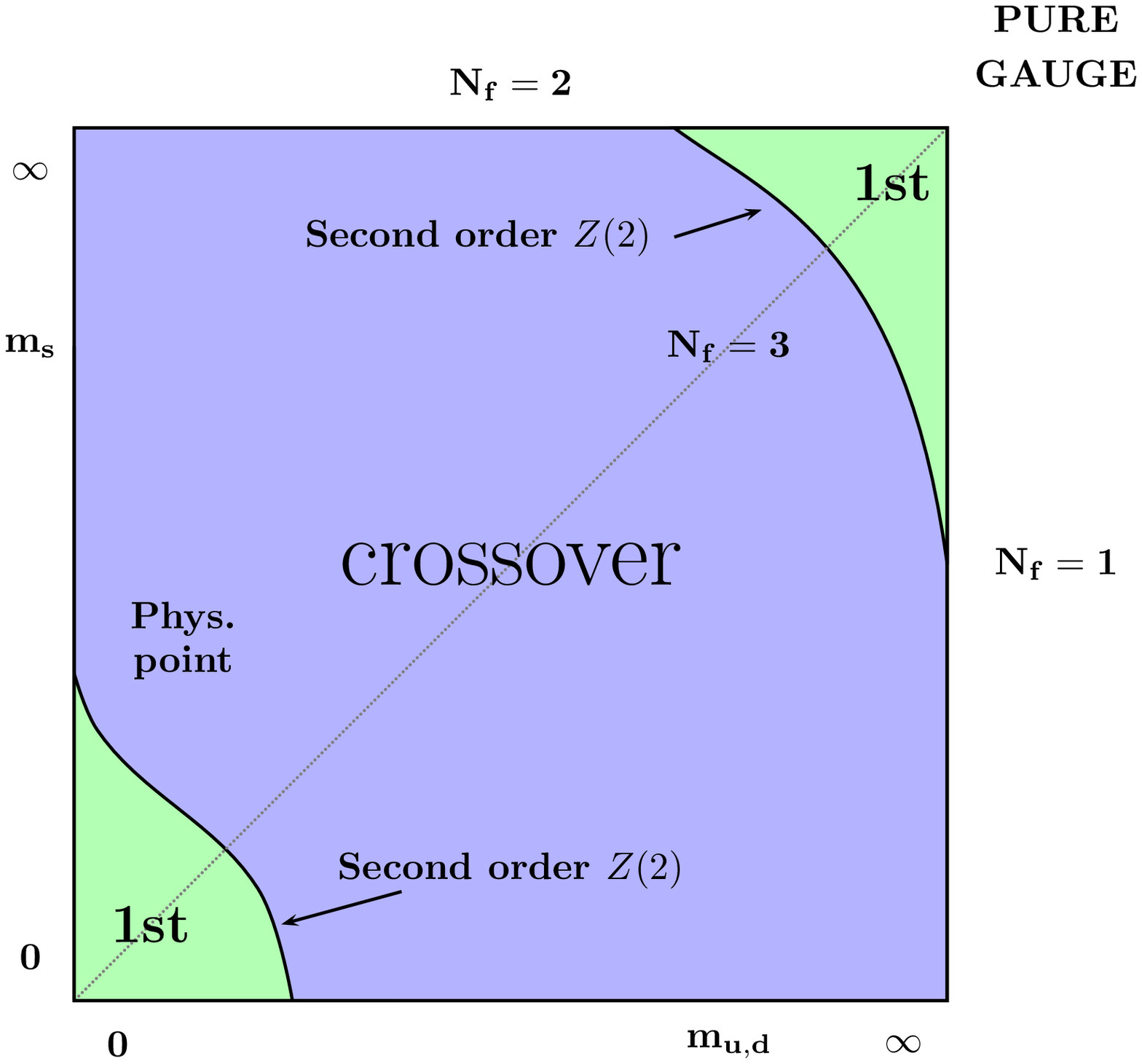}
}\label{fig_phd}
\vspace*{-5.5cm}
\caption{Schematic overview of the $N_f=2+1$ QCD phase diagram. Here we compute the critical line for heavy quarks.}
\end{center}
\end{figure}

After a successful test of our method for $SU(2)$ Yang-Mills theory, we study the case 
of $SU(3)$ QCD with Wilson quarks by a combined strong coupling and hopping parameter 
expansion, which converges 
for sufficiently heavy quarks. In the infinite quark mass limit, QCD is known to 
display a first order phase transition which weakens as quark masses are lowered, 
see \fig\ref{fig_phd}.
The location of the critical quark mass, where the transition disappears at a second order critical point, has been studied for $N_f=1$ by a combination of 
numerical simulations of QCD and the 3d 3-state 
Potts model \cite{potts,ks}, which is the appropriate effective theory for the critical 
transition. These works also established the 3d Ising universality of the boundary line.
Our methods allow for a determination of the critical quark mass for $N_f=1,2,3$, 
as well as the dependence of the critical
quark mass on quark chemical potential. We find that the critical mass grows with 
increasing chemical potential, in accord with a Monte Carlo study in the 
Potts model with finite chemical potential \cite{kfkt}. 
However, our result is the first based on full QCD beyond mean field theory.

\section{Phase transitions from strong coupling series}

\subsection{Notation and formalism}

We work on a $(3+1)$-dimensional hypercubic lattice with lattice spacing $a$ and 
infinite spatial volume.
The temporal lattice extent $N_\tau$ is kept finite, 
which in combination with (anti-)periodic boundary conditions for (fermionic) bosonic fields 
generates a nonvanishing physical temperature,
\begin{eqnarray}
T=\frac{1}{N_{\tau}a}\;.
\end{eqnarray}
Using Wilson's gauge action for $SU(N)$, the partition function reads
\begin{eqnarray}
Z=\int \left[dU\right]\,\exp\left(-S_g\right)=\int \left[dU\right]\,\prod_p\exp\left[\frac{\beta}{2N}\,\Big(\mathrm{tr}\,U_p+\mathrm{tr}\,U^{\dagger}_p\Big)\right].
\end{eqnarray}
In order to locate the phase transition we consider 
the Polyakov loop susceptibility 
\begin{eqnarray}
\chi_L&=&V\Big(\langle L^2\rangle-\langle L\rangle^2\Big),
\end{eqnarray}
where we have defined the Polyakov loop $L_{\vec{x}}$ and its spatial average $L$ as
\begin{eqnarray}
L_{\vec{x}}&=&\mathrm{tr}\,W_{\vec{x}}=\mathrm{tr}\,\prod_{\tau=1}^{N_{\tau}}U_0(\vec{x},\tau),\\
\nonumber\\
L&=&\frac{1}{V}\sum_{\vec{x}}\Big(L_{\vec{x}}+L^{\dagger}_{\vec{x}}\Big)\qquad SU(N\geq3)\nonumber\\
L&=&\frac{1}{V}\sum_{\vec{x}}L_{\vec{x}}\qquad\qquad\qquad SU(2).
\end{eqnarray}
If we couple the Polyakov loop to an external source $J$ in the action\footnote{
For $SU(3)$ we have chosen $L$ as the real part of the Polyakov loop, 
so we get a real action with only one real source $J$}, 
\begin{eqnarray}
-S(J)=\frac{\beta}{2N}\,\sum_p\Big(\mathrm{tr}\,U_p+\mathrm{tr}\,U^{\dagger}_p\Big)+J\sum_{\vec{x}}\Big(L_{\vec{x}}+L^{\dagger}_{\vec{x}}\Big),
\end{eqnarray}
we can express the susceptibility as
\begin{eqnarray}
\chi_L=\frac{1}{V}\,\frac{\partial^2}{\partial J^2}\,\mathrm{ln}\,Z(J)\bigg\vert_{J=0}.
\label{eq_pls}\end{eqnarray}

In order to obtain a strong coupling series for \eq(\ref{eq_pls}), 
we expand the partition function
\begin{eqnarray}
Z(J)=\int \left[dU\right]\left[\prod_p\exp\left(\frac{\beta}{2N}\,\Big(\mathrm{tr}\,U_p+\mathrm{tr}\,U^{\dagger}_p\Big)\right)\right]\left[\prod_{\vec{x}}\exp\bigg(J\,\Big(\mathrm{tr}W_{\vec{x}}+\mathrm{tr}W^{\dagger}_{\vec{x}}\Big)\bigg)\right]
\end{eqnarray}
in terms of characters
\begin{eqnarray}
\exp\left(\frac{\beta}{2N}\,\Big(\mathrm{tr}\,U_p+\mathrm{tr}\,U^{\dagger}_p\Big)\right)&=&\left[1+\sum_{r\neq0}d_ra_r\left(\beta\right)\chi_r(U_p)\right]\\
\nonumber\\
\exp\bigg(J\,\Big(\mathrm{tr}W_{\vec{x}}+\mathrm{tr}W^{\dagger}_{\vec{x}}\Big)\bigg)&=&c_0(J)\bigg[1+\sum_{r\neq0}b_r(J)\chi_r(W_{\vec{x}})\bigg].
\end{eqnarray}
We have neglected the prefactor $c_0(\beta)$ of the trivial representation because 
it does not depend on $J$ and vanishes in \eq (\ref{eq_pls}). 
As usual in strong coupling expansions, 
we will use the coefficient $u=a_{f}$ of the fundamental representation 
as expansion parameter.

In the case of $SU(2)$ we have no complex conjugate representations and 
the partition function reads
\begin{eqnarray}
Z(J)=\int \left[dU\right]\left[\prod_p\exp\left(\frac{\beta}{2}\,\mathrm{tr}\,U_p\right)\right]\left[\prod_{\vec{x}}\exp\bigg(J\,\mathrm{tr}W_{\vec{x}}\bigg)\right].
\end{eqnarray}
Furthermore, there are closed form expressions for the expansion coefficients 
in this case,
\begin{eqnarray}
a_j\left(\beta\right)&=&\frac{I_{2j+1}(\beta)}{I_1(\beta)},\\
\nonumber\\\
b_j(J)&=&\frac{d_jI_{2j+1}(2J)}{I_1(2J)}\label{eq_h},\\
\nonumber\\
c_0(J)&=&\frac{I_1(2J)}{J}.
\end{eqnarray}

Applying a cluster expansion as described in \cite{mm},
the logarithm of the partition function
can be represented as a sum of graphs $\Phi(C)$
\begin{eqnarray}
\frac{1}{V}\mathrm{ln}\,Z(J)=\mathrm{ln}\,c_0(J)+\sum_{C}\Phi(C).
\end{eqnarray}
The sum is over all clusters $C$ of connected polymers, see \cite{La} for details. 
The leading order, i.e.~the strong coupling limit $\beta=0$, 
is obtained by neglecting all graphs giving the trivial result
\begin{eqnarray}
\chi_L&=&\frac{\partial^2}{\partial J^2}\,\mathrm{ln}\,c_0(J)\bigg\vert_{J=0}=1+{\cal{O}}(u^{N_{\tau}})\qquad SU(2)\nn\\
\chi_L&=&\frac{\partial^2}{\partial J^2}\,\mathrm{ln}\,c_0(J)\bigg\vert_{J=0}=2+{\cal{O}}(u^{N_{\tau}})\qquad SU(N\geq3).
\end{eqnarray}
In case of $SU(N\geq3)$ the factor of $2$ accounts for both fundamental representations.

\begin{figure}
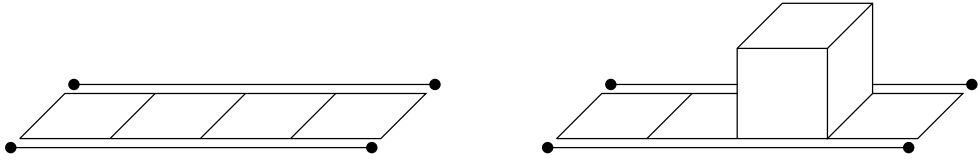

\hspace{1.5cm}
\begin{minipage}{7cm}
\scalebox{0.6}{
\psline(0,0)(8,0)(9,1)(1,1)(0,0)
\psline(2,0)(3,1)
\psline(4,0)(5,1)
\psline(6,0)(7,1)
\psline(-0.2,-0.2)(7.8,-0.2)\psdot[dotscale=2 2](-0.2,-0.2)\psdot[dotscale=2 2](7.8,-0.2)
\psline(1.2,1.2)(9.2,1.2)\psdot[dotscale=2 2](1.2,1.2)\psdot[dotscale=2 2](9.2,1.2)
}
\end{minipage}
\begin{minipage}{6cm}
\scalebox{0.6}{
\psline(-0.2,-0.2)(7.8,-0.2)\psdot[dotscale=2 2](-0.2,-0.2)\psdot[dotscale=2 2](7.8,-0.2)
\psline(0,0)(8,0)(9,1)(7,1)(7,3)(5,3)(4,2)(4,1)(1,1)(0,0)
\psline(2,0)(3,1)
\psline(7,1)(6,0)(6,2)(7,3)
\psline(4,0)(4,1)
\psline(4,2)(6,2)
\psline(1.2,1.2)(4,1.2)
\psline(7,1.2)(9.2,1.2)\psdot[dotscale=2 2](1.2,1.2)\psdot[dotscale=2 2](9.2,1.2)
}
\end{minipage}
\caption{Examples for $N_{\tau}=4$. Left: Two Polyakov loops and $N_{\tau}$ plaquettes wind around the temporal dimension. Right: The first correction with additional plaquettes.}
\label{fig_lo}
\end{figure}

The first graph with a non-trivial $u$-dependence is shown in \fig\ref{fig_lo}, 
together with the leading correction graph. 
The left and the right boundaries of these graphs are meant to be identified due to the 
periodic boundary conditions. To calculate the contribution of these graphs, we employ the 
group integration formula
\begin{eqnarray}
\int dU\chi_r(U)\chi_s(U^{\dagger})=\frac{\delta_{rs}}{d_{r}},
\end{eqnarray}
and get
\begin{eqnarray}
\Phi_0&=&3u^{N_{\tau}}\left(b_{1/2}(J)\right)^2\qquad SU(2),\nn\\
\Phi_0&=&6u^{N_{\tau}}\left(b_{f}(J)\right)^2\qquad SU(N\geq3),
\end{eqnarray}
for the leading non-trivial order and
\begin{eqnarray}
\Phi_1&=&12N_{\tau}u^{N_{\tau}+4}\left(b_{1/2}(J)\right)^2\qquad SU(2),\nn\\
\Phi_1&=&24N_{\tau}u^{N_{\tau}+4}\left(b_{f}(J)\right)^2\qquad SU(N\geq3).
\end{eqnarray}
for the first correction.

\subsection{Graphical expansion}

From \eq(\ref{eq_pls}) it is obvious that we solely have to take into account graphs 
which contribute to order $J^2$. This means that 
only graphs with two Polyakov loops in the fundamental or one loop in the adjoint 
representation are allowed.
For the first possibility, the loops have to be on different lattice sites.
The generation of contributing graphs is not uniform and we distinguish 
between small, intermediate and large $N_{\tau}$. 

\paragraph{Large $N_{\tau}$:}

Large $N_{\tau}$ receive only contributions from nearest-neighbour Polyakov loops as 
shown in \fig\ref{fig_lo}, and corrections from adding plaquettes. Of course, 
this statement is only true for large enough $N_{\tau}$ if we calculate to some 
fixed order in $u$.

\paragraph{Small $N_{\tau}$:}

\begin{figure}[t]
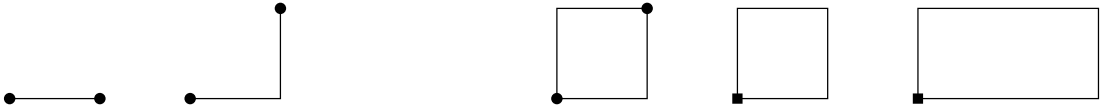

\hspace{0.5cm}
\begin{minipage}{6cm}
\scalebox{0.6}{
\psline(0,0)(2,0)\psdot[dotscale=2 2](0,0)\psdot[dotscale=2 2](2,0)
\psline(4,0)(6,0)(6,2)\psdot[dotscale=2 2](4,0)\psdot[dotscale=2 2](6,2)
}
\end{minipage}
\hspace{1cm}
\begin{minipage}{6cm}
\scalebox{0.6}{
\psline(0,0)(2,0)(2,2)(0,2)(0,0)\psdot[dotscale=2 2](0,0)\psdot[dotscale=2 2](2,2)
\psline(4,0)(6,0)(6,2)(4,2)(4,0)
\psdot[dotstyle=square*,dotscale=2 2](4,0)
\psline(8,0)(12,0)(12,2)(8,2)(8,0)
\psdot[dotstyle=square*,dotscale=2 2](8,0)
}
\end{minipage}
\caption{Left: Self avoiding walks with two fundamental Polyakov loops.  Right: Self avoiding polygons with one adjoint or two fundamental Polyakov loop.}
\label{fig_saw}
\end{figure}

The smallest possible $N_{\tau}$ is 1. Typical graphs are shown in \fig\ref{fig_saw}. 
These graphs are meant to be spatial projections of graphs like in \fig\ref{fig_lo} 
(left). In higher orders we get contributions from 
additional spatial plaquettes, e.g.~by filling the cross-section of the 
self-avoiding polygons, but these contributions are small compared to the increasing 
number of self-avoiding walks.

\paragraph{Intermediate $N_{\tau}$:}

\begin{figure}[t]
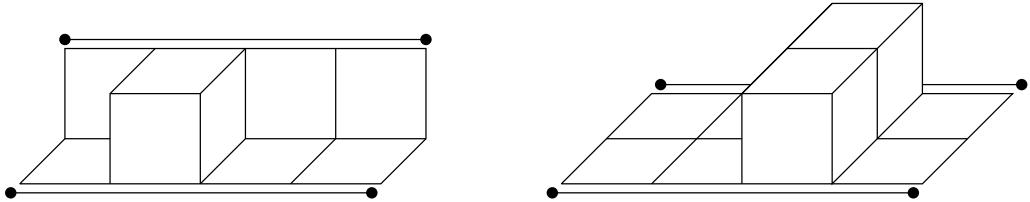

\vspace{3.5cm}
\hspace{1.5cm}
\scalebox{0.6}{
\psline(0,0)(8,0)(9,1)(9,3)(1,3)(1,1)(0,0)
\psline(3,3)(2,2)(2,0)
\psline(5,3)(4,2)(4,0)
\psline(5,3)(5,1)(4,0)
\psline(7,3)(7,1)(6,0)
\psline(1,1)(2,1)
\psline(5,1)(9,1)
\psline(2,2)(4,2)
\psline(-0.2,-0.2)(7.8,-0.2)\psdot[dotscale=2 2](-0.2,-0.2)\psdot[dotscale=2 2](7.8,-0.2)
\psline(1,3.2)(9,3.2)\psdot[dotscale=2 2](1,3.2)\psdot[dotscale=2 2](9,3.2)
\psline(12,0)(20,0)(22,2)(20,2)(20,4)(18,4)(16,2)(14,2)(12,0)
\psline(20,2)(18,0)(18,2)(20,4)
\psline(16,0)(16,2)(18,4)
\psline(17,3)(19,3)(19,1)(21,1)
\psline(16,2)(18,2)
\psline(14,0)(16,2)
\psline(13,1)(16,1)
\psline(11.8,-0.2)(19.8,-0.2)\psdot[dotscale=2 2](11.8,-0.2)\psdot[dotscale=2 2](19.8,-0.2)
\psline(14.2,2.2)(16.2,2.2)\psline(20,2.2)(22.2,2.2)\psdot[dotscale=2 2](14.2,2.2)\psdot[dotscale=2 2](22.2,2.2)
}
\caption{Examples of corrections to self avoiding walks of length $L=2$ and $N_{\tau}=4$.}
\label{fig_corr}
\end{figure}

For intermediate $N_{\tau}$ (=2,3,4...) we have to take into account graphs of both types. 
There are also some other corrections as shown in \fig \ref{fig_corr}. Thus these 
$N_{\tau}$ are the most labour-intensive ones.

\subsection{Series analysis and phase transitions}

Strong coupling expansions have a finite radius of convergence. Expanding about
$\beta=0$, a true deconfinement phase transition at some critical value of $\beta_c$
clearly represents an upper bound on the convergence radius, i.e.~strong coupling
analyses are limited to the confined region. Nevertheless, knowledge of the series
coefficients allows us to estimate the location of a singularity along the 
real $\beta$-axis. 
Our analysis is well suited to detect second order phase transitions and exploits 
the fact that the Polyakov loop susceptibility in this case
diverges with a critical exponent.
Near a critical coupling the Polyakov loop susceptibility and its logarithmic 
derivative behave like
\begin{eqnarray}
\chi_L\sim\frac{1}{(u_c-u)^{\gamma}}\;,\qquad\qquad
D_{\chi}(u)\equiv\frac{d}{du}\mathrm{ln}\,(\chi_L)\sim\frac{\gamma}{(u_c-u)}.
\label{eq_dlog}\end{eqnarray}
From our series expansions we know $D_{\chi}(u)$ as a polynomial in $u$ and 
can model its pole-like singularity by Pad\'e 
approximants 
\begin{eqnarray}[L,M](u)\equiv 
\frac{a_0+a_1u+\dots +a_Lu^L}{1+b_1u+\dots+b_Mu^M}.
\end{eqnarray}
In order to uniquely determine the coefficients $a_i,b_i$, it is necessary to have $L+M\leq N$, if $N$ represents the highest available order of the expansion. In this way a $\left[L,M\right]$ approximant is correct up to but not including ${\cal{O}}(u^{L+M+1})$ and larger approximants represent more expansion coefficients than smaller ones.
In particular, the critical coupling $u_c$ is given as the real positive zero of the
denominator closest to the origin, the critical exponent $\gamma$ is obtained
from the corresponding residuum.

The analysis can be made more powerful if either independent results for the
critical couplings are available, or the universality class of the transition is known.
In the first case, it is possible to get better estimates for the critical exponents via
\begin{eqnarray}
(u_c-u)D_{\chi}(u)=\gamma+{\cal{O}}(u_c-u),
\end{eqnarray}
which is the more precise the better we know $u_c$. In order to do so, we calculate Pad\'e approximants to the series expansion of $(u_c-u)D_{\chi}(u)$ and evaluate them at the known critical coupling $u_c$.
In the same way we can use a known value $\gamma$ to get more accurate
estimates for the critical coupling
\begin{eqnarray}
\left(\chi_L\right)^{\frac{1}{\gamma}}\sim\frac{1}{(u_c-u)}\;.
\end{eqnarray}
Here we calculate Pad\'e approximants to $\left(\chi_L\right)^{\frac{1}{\gamma}}$ and solve for zeros of the denominator, as this quantity has a simple pole at the critical coupling.
For a more detailed discussion of these topics, see \cite{gutt}.

\section{$SU(2)$ Yang-Mills}

\subsection{Results for the series}

We first apply our analysis method to $SU(2)$ pure gauge theory, where we
have reasonably long series for $N_\tau=1-4$ and where accurate Monte Carlo data
are available for comparison.
We obtain the following strong coupling series for $\chi_L(N_{\tau},u)$:
\begin{eqnarray}
\chi_L(1,u)&=&1+6\,u+30\,{u}^{2}+150\,{u}^{3}+738\,{u}^{4}+3622\,{u}^{5}+{\frac {
52982}{3}}{u}^{6}+\nonumber\\
&&+\,{\frac {773434}{9}}{u}^{7}+{\frac {11239612}{27}}{u}
^{8}+{\cal{O}}\left( {u}^{9} \right),\nonumber\\
\chi_L(2,u)&=&1+6\,{u}^{2}+30\,{u}^{4}+222\,{u}^{6}+1218\,{u}^{8}+{\frac {24602}{3}}{u}^{10}+{\cal{O}} \left( {u}^{12} \right),\nonumber\\
\chi_L(3,u)&=&1+6\,{u}^{3}+30\,{u}^{6}+72\,{u}^{7}+72\,{u}^{8}+78\,{u}^{9}+576\,{u}
^{10}+1776\,{u}^{11}+\nonumber
\\
&&+\,1770\,{u}^{12}+{\cal{O}}\left( {u}^{13} \right),\nonumber\\
\chi_L(4,u)&=&1+6\,{u}^{4}+126\,{u}^{8}+48\,{u}^{10}+2830\,{u}^{12}+{\frac {91808}{
135}}{u}^{14}+{\cal{O}}\left({u}^{16}
 \right).
\end{eqnarray}
$N_{\tau}=1$ corresponds to the largest lattice spacing and thus to the largest bare coup\-ling at the deconfinement transition. Hence, our series shows the best 
convergence behaviour in this case. 

\subsection{The critical parameters}

We consider only Pad\'es $[L,M]$ with $L,M>2$ 
in order to have large enough polynomials in both numerator and denominator. 
So-called defective approximants with an adjacent zero-pole pair,
indicated by a small residuum (we chose $Res<0.003$ to be defective), 
are also ignored. By doing so we obtain estimates for the critical parameters as shown in 
Table \ref{tab_unb}. In the last line we have averaged over the estimates from different
Pad\'es in order to quantify the systematic error associated with the choice of a 
particular approximant. Note that the quoted error is estimated as $(\beta_c^{max}-\beta_c^{min})/2$, and similarly
for the exponent.
It is only due to the scatter in the singularity structure 
of different Pad\'e approximants and does not include 
the error from the truncation of the series, i.e.~it likely underestimates the true error.

\begin{table}[t]
\centering
\begin{tabular}{|c|c|c|c|}
\hline
 Pad\'e           & $u_c$     & $\beta_c$ & $\gamma$
\\\hline\hline
$\left[5,2\right]$& $0.21055$ & $0.86825$ & $1.167$ \\
\hline
$\left[3,3\right]$& $0.20967$ & $0.86439$ & $1.138$ \\
$\left[4,2\right]$& $0.20957$ & $0.86396$ & $1.146$ \\
$\left[2,4\right]$& $0.20987$ & $0.86527$ & $1.135$ \\
\hline
$\left[3,2\right]$& $0.20927$ & $0.86264$ & $1.126$ \\
\hline
$\left[2,2\right]$& $0.20820$ & $0.85796$ & $1.102$ \\
\hline\hline
Mean & $0.2095(12)$ & $0.864(5)$ & $1.14(3)$\\
\hline
\end{tabular}\caption{Critical coupling and exponent for  
$N_{\tau}=1$, estimated from different Pad\'e approximants.}
\label{tab_unb}
\end{table}

Our result for a $N_\tau=1$ lattice then is $\beta_c=0.864(5), \gamma=1.14(3)$.
There exist different values for the critical coupling from Monte Carlo simulations 
in the literature: $\beta_c=0.8730(2)$ from \cite{BA} and the more recent 
$\beta_c=0.85997(10)$ or $0.86226(6)$ from \cite{Ve}. 
At first sight our results appear to favour one of the latter, but with a critical exponent 
deviating about 10\% from universality. 
While the Pad\'e approximants accumulate a pole and thus
definitely predict a second order phase transition, the residuum is somewhat 
below the value $\gamma_I=1.2373(2)$ \cite{Ca} for a 3d Ising transition.
Note however the upward trend in the critical coupling as well as in the exponent with 
increasing order $L+M$ of the approximants in Table \ref{tab_unb}. 
This indicates that the results are not yet fully stable and the exponent should
reach the Ising value with longer series.

It is apparent that we have gained considerable accuracy
compared to previous work 
using the free energy and its derivatives as observables \cite{La}, which gives
$\beta_c=0.92(15), \alpha=0.063(38)$
($\alpha=0.1096(5)$ for 3d Ising \cite{Ca}). The reason for this improvement is twofold: 
the Polyakov loop susceptibility permits an easier evaluation of more coefficents,
e.g.~by featuring both even and odd powers of $u$, 
and the series itself comes with only positive coefficients and\begin{large}                                                               \end{large} is 
better behaved than that for the free energy. 

It is now interesting to explore how one can combine Monte Carlo and series results. 
Thus we consider biased estimates, which should be more accurate. 
Using $\gamma_I=1.237$, we get the results shown in 
Table \ref{tab_crit}.
\begin{table}
\begin{minipage}{9cm}
\centering
\begin{tabular}{|c|c|c|c|}
\hline
 Pad\'e           & $u_c$     & $\beta_c$ 
\\\hline\hline 
$\left[6,2\right]$& $0.21221$ & $0.87553$\\
\hline
$\left[4,3\right]$& $0.21159$ & $0.87281$\\
$\left[2,5\right]$& $0.21138$ & $0.87189$\\ 
\hline
$\left[3,3\right]$& $0.21229$ & $0.87588$\\
$\left[2,4\right]$& $0.21238$ & $0.87628$\\
$\left[4,2\right]$& $0.21279$ & $0.87808$\\
\hline
$\left[3,2\right]$& $0.20986$ & $0.86523$\\ 
$\left[2,3\right]$& $0.21464$ & $0.88621$\\
\hline
$\left[2,2\right]$& $0.21495$ & $0.88757$\\
\hline
\end{tabular}
\end{minipage}
\begin{minipage}{5cm}
\centering
\begin{tabular}{|c|c|c|c|}
\hline
 Pad\'e           & $\gamma_1$ &  $\gamma_2$ 
\\\hline\hline 
$\left[3,4\right]$& $1.1250$ &  $1.2378$ \\
$\left[4,3\right]$& $1.1244$ &  $1.2331$ \\ 
$\left[2,5\right]$& $1.1246$ &  $1.2157$ \\
$\left[5,2\right]$& $1.1244$ &  $1.2208$ \\
\hline
$\left[3,3\right]$& $1.1225$ &  $1.2579$ \\
$\left[2,4\right]$& $1.1236$ &  $1.2661$ \\
$\left[4,2\right]$& $1.1244$ &  $1.2950$ \\
\hline
$\left[3,2\right]$& $1.1240$ &  $1.2308$ \\ 
$\left[2,3\right]$& $1.1238$ &  $1.2215$ \\
\hline
\end{tabular}
\end{minipage}
\caption{Biased critical couplings and exponents for $N_{\tau}=1$.}
\label{tab_crit}
\end{table}
We calculate the average of  $\beta_c$ to be $\overline{\beta_c}=0.877(11)$ using 
all Pad\'es and $\overline{\beta_c}=0.875(3)$ using only the three highest orders 
which behave more smoothly. 
Despite the fact that the total error is underestimated,
both estimates are consistent with the Monte Carlo result of \cite{BA}.

In order to obtain the biased critical exponent, we used the values 
$\beta_1=0.86226$ and $\beta_2=0.873$. 
The former gives a mean critical exponent of $\overline{\gamma_1}=1.124(1)$ and the 
latter $\overline{\gamma_2}=1.24(4)$. Although the first result is much more stable, 
it is the second one which is consistent with universality. Hence we conclude that 
it is the value $\beta_c=0.8730(2)$ of \cite{BA}, which is supported by our series 
expansions.

For intermediate $N_{\tau}$ our results become less precise the larger 
we choose $N_{\tau}$. This is to be expected since $\beta_c$ grows on finer lattices and 
we are leaving the strong coupling regime. 
Thus we only give our biased estimates, 
using the Monte Carlo results 
$\beta_c(N_{\tau}=2)=1.87348, \beta_c(N_{\tau}=3)=2.1768, 
\beta_c(N_{\tau}=4)=2.2993$ \cite{Fi},\cite{Ve} and $\gamma_I=1.237$.
We summarise our results in Table \ref{tab_nt} and observe that
the predicted quantities are fully consistent with Monte Carlo 
results and universality. 

\begin{table}[ht]
\begin{minipage}{15cm}
\centering
\begin{tabular}{|c|c|c|c|c|c|c|}
\hline
 $N_{\tau}$ & $\gamma_I$ & $\overline{\gamma}$ & No. of Pad\'es & $\beta_c^{MC}$ & $\overline{\beta_c}$ & No. of Pad\'es
\\\hline\hline
$2$         & $1.237$    & $1.21(2)$           & $5$     & $1.87348(2)$   & $1.87(1)$ & 6 \\ 
\hline
$3$         & $1.237$    & $1.29(18)$            & $2$     & $2.1768(30)$   & $2.13(4)$  & 6 \\
\hline
$4$         & $1.237$    & $1.22(20)$            & $3$     & $2.2993(3)$    & $2.23(11)$  & 7 \\ 
\hline
\end{tabular}
\caption{Comparison of our findings, $\overline{\gamma},\overline{\beta_c}$, with
the values from universality and simulations.}
\label{tab_nt}
\end{minipage}
\end{table}

\section{$SU(3)$ and QCD}

For $SU(3)$ Yang-Mills theory, there is a first order phase transition, i.e.~the correlation 
length remains finite even at the critical temperature, spoiling our analysis method which requires scaling behaviour. 
We therefore introduce heavy dynamical quarks, 
which have a critical mass $m_c$ where the transition turns second order,
cf.~\fig\ref{fig_phd}. 
It is this point which we now try to locate.

\subsection{Combined strong coupling and hopping expansion}

We introduce dynamical quarks in leading order hopping parameter expansion (see \cite{mm} for details), where the quark part of the action reads
\begin{eqnarray}
S_q=\sum_l\frac{\kappa^l}{l}\mathrm{tr}\,M[U]^l, \qquad\qquad
\kappa=\frac{1}{2m+8},
\label{eq_quark_action}
\end{eqnarray}
$m$ is the quark mass and $M[U]$ the quark hopping matrix
\begin{eqnarray}
M[U]_{yx}=\sum_{\mu}\delta_{y,x+\hat{\mu}}(1+\gamma_{\mu})U_{x\mu}.
\end{eqnarray}
The sum in \eq(\ref{eq_quark_action}) extends over all closed paths on the lattice. For small temporal lattice sizes and finite temperature the leading order hopping expansion term is just the Polyakov loop $\mathrm{tr}\,W_{\vec{x}}$.
Chemical potential is introduced in the usual way as factors $\exp(\pm\mu)$ to the temporal 
link variables \cite{hasenfratz}. 
The effective quark part of the action for small temporal lattice extents then reads
\begin{eqnarray}
-S_q^{eff}=\sum_{\vec{x}}\Big[h( \kappa)e^{\mu}\,\mathrm{tr}\,W_{\vec{x}}+h(\kappa)e^{-\mu}\,
\mathrm{tr}\,W^{\dagger}_{\vec{x}}\Big],
\end{eqnarray}
where the relative minus sign compared to \eq(\ref{eq_quark_action}) is due to the antiperiodic boundary conditions for fermions.
The parameter $h(\kappa)$ depends on the hopping parameter $\kappa$ and the number of degenerate quark flavours $N_f$ via
\begin{eqnarray}
h(\kappa)=2N_f(2\kappa)^{N_{\tau}}.
\label{eq_hopping}
\end{eqnarray}
For $\mu\neq 0$ the action is complex.

\begin{figure}[t]
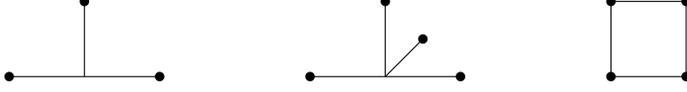

\hspace{1cm}
\scalebox{0.5}{
\psline(0,0)(4,0)
\psline(2,0)(2,2)\psdot[dotscale=2 2](0,0)\psdot[dotscale=2 2](2,2)
\psdot[dotscale=2 2](4,0)
\psline(8,0)(12,0)
\psline(10,0)(10,2)
\psline(10,0)(11,1)
\psdot[dotscale=2 2](8,0)\psdot[dotscale=2 2](10,2)
\psdot[dotscale=2 2](12,0)\psdot[dotscale=2 2](11,1)
\psline(16,0)(18,0)(18,2)(16,2)(16,0)\psdot[dotscale=2 2](16,2)\psdot[dotscale=2 2](18,2)\psdot[dotscale=2 2](16,0)\psdot[dotscale=2 2](18,0)
}
\caption{Examples of terms with a larger number of Polyakov loop source terms.}
\label{fig_qcd}
\end{figure}

Introducing Polyakov loop source terms $J$ and putting everything together, we
obtain the following partition function
\begin{eqnarray}
Z&=&\int\left[dU\right]\exp\bigg[\frac{\beta}{6}\sum_p\Big(\mathrm{tr}U_p+\mathrm{tr}U_p^{\dagger}\Big)\nonumber\\
&+&\sum_{\vec{x}}\Big\lbrace  \big[he^{\mu}+J\big]\mathrm{tr}\,W_{\vec{x}}+\big[he^{-\mu}+J\big]\mathrm{tr}\,W^{\dagger}_{\vec{x}}\Big\rbrace\bigg].
\end{eqnarray}
Now we can proceed in the same way as in case of $SU(2)$ and rewrite the partition function with two different character expansions 
and omit the factor $c_0(\beta)$
\begin{eqnarray}
Z&=&\int\left[dU\right]\prod_{p}\left[1+\sum_{r\neq0}d_ra_r(\beta)\chi_r(U_p)\right]\nonumber\\
&\times&\prod_{\vec{x}}c_0(h,\mu,J)\left[1+\sum_{r\neq0}b_r(h,\mu,J)\chi_r(W_{\vec{x}})\right].
\end{eqnarray}
Note that for non-vanishing chemical potential $\mu$ the 
expansion parameters $b_r\big(h,\mu,J\big)$ are different for complex 
conjugate representations $r$ and $\bar{r}$ and related via 
\begin{eqnarray}
b_r\big(h,\mu,J\big)=b_{\bar{r}}\big(h,-\mu,J\big).
\end{eqnarray}
The expansion coefficients itself can be expressed as series expansions, e.g.
\begin{eqnarray}
u\equiv a_f(\beta)&=&\frac{1}{18}\,\beta+\ldots,\nn\\
c_0(h,\mu,J)&=&1+\big(he^{\mu}+J\big)\big(he^{-\mu}+J\big)+\ldots,\nn\\
b_f(h,\mu,J)&=&he^{\mu}+J+\ldots.
\end{eqnarray}
In order to get the proper series expansion we now have to draw all possible diagrams 
to a given order in $u$ and the number of Polyakov loops $l$. In contrast to our $SU(2)$ calculation we 
have to take into account not only the graphs with two Polyakov loop source terms, 
but all graphs with contributions of the order $J^2h^m$, since these give finite results 
after differentiating twice with respect to $J$ and setting $J=0$. Some examples of graphs for the case $N_{\tau}=1$ are given in \fig\ref{fig_qcd}.
An important fact is that for a given order in $u$, there is only a finite number 
of graphs. For the order $u^n$, we can have only graphs fulfilling $l\leq2n$, 
as inspection of the series shows. 
Additional terms in the hopping expansion will modify the $b_r$'s and rapidly 
increase the number of relevant graphs.

\subsection{Result for the series}

We have derived the series expansion of the Polyakov loop susceptibility up to orders $u^nh^m$, with $n+m\leq6$. The result for the $N_{\tau}=1$ series, arranged in increasing orders of $u$, reads
\begin{eqnarray}
\chi_L(u,h)&=&\bigg[1+ch+\left(-\frac{4}{3}c^3+\frac{1}{2}c\right)h^3+\left(-\frac{5}{3}c^4+\frac{4}{3}c^2-\frac{7}{24}\right)h^4\nonumber\\
&&+\left(\frac{2}{15}c^5+\frac{1}{3}c^3-\frac{1}{8}c\right)h^5+\left(\frac{28}{15}c^6-\frac{7}{5}c^4-\frac{7}{120}c^2+\frac{119}{720}\right)h^6\bigg]\nonumber\\
&&+\bigg[6+18ch+\left(6c^2+3\right)h^2+\left(-40c^3+15c\right)h^3\nonumber\\
&&+\left(-90c^4+66c^2-\frac{69}{4}\right)h^4+\left(-\frac{32}{5}c^5-8c^3+6c\right)h^5\bigg]u\nonumber\\
&&+\bigg[30+180ch+\left(144c^2+72\right)h^2+\left(-760c^3+285c\right)h^3\nonumber\\
&&+\left(-\frac{5985}{2}c^4+\frac{8985}{4}c^2-\frac{4485}{8}\right)h^4\bigg]u^2\nonumber\\
&&+\bigg[150+1470ch+\left(\frac{4113}{2}c^2+\frac{4113}{4}\right)h^2\nonumber\\
&&+\left(-6856c^3+2571c\right)h^3\bigg]u^3\nonumber\\
&&+\bigg[786+10752ch+\left(\frac{1131747}{32}c^2+\frac{1088547}{64}\right)\bigg]u^4\nonumber\\
&&+\bigg[4011+73521ch\bigg]u^5+\frac{152247}{8}u^6,
\label{eq_u_h_mu}\end{eqnarray}
where we used the abbreviation $c\equiv\cosh(\mu)$. Since the only $\mu$-dependence 
appears in $\cosh(\mu)$ terms, one can immediately see that the 
Polyakov loop susceptibility is invariant under $\mu\leftrightarrow-\mu$ as it should be
according to the charge conjugation symmetry of QCD.

\subsection{Critical point for $\mu=0$}

\begin{figure}
\vspace*{2cm}
\hspace*{1cm}
\begin{minipage}{3cm}
\scalebox{1}{
\psline{->}(0,0)(4,0)\psline{->}(0,0)(0,4)
\psbezier(3,0)(2.75,0.5)(2.25,1)(1.5,1.5)\psdot(1.5,1.5)
\psline[linestyle=dotted]{->}(0,0)(2.5,2.5)
\rput(3,-0.5){u}\rput(-0.5,3){h}\rput(2.5,2){t}
\rput(3.5,1){first order line}
\rput(2,3){second order point}
\psline[linewidth=0.01]{->}(1.5,2.8)(1.5,1.8)}
\end{minipage}
\hspace*{3cm}
\begin{minipage}{4cm}
\vspace*{-3cm}
\scalebox{0.3}{
\includegraphics[angle=-90]{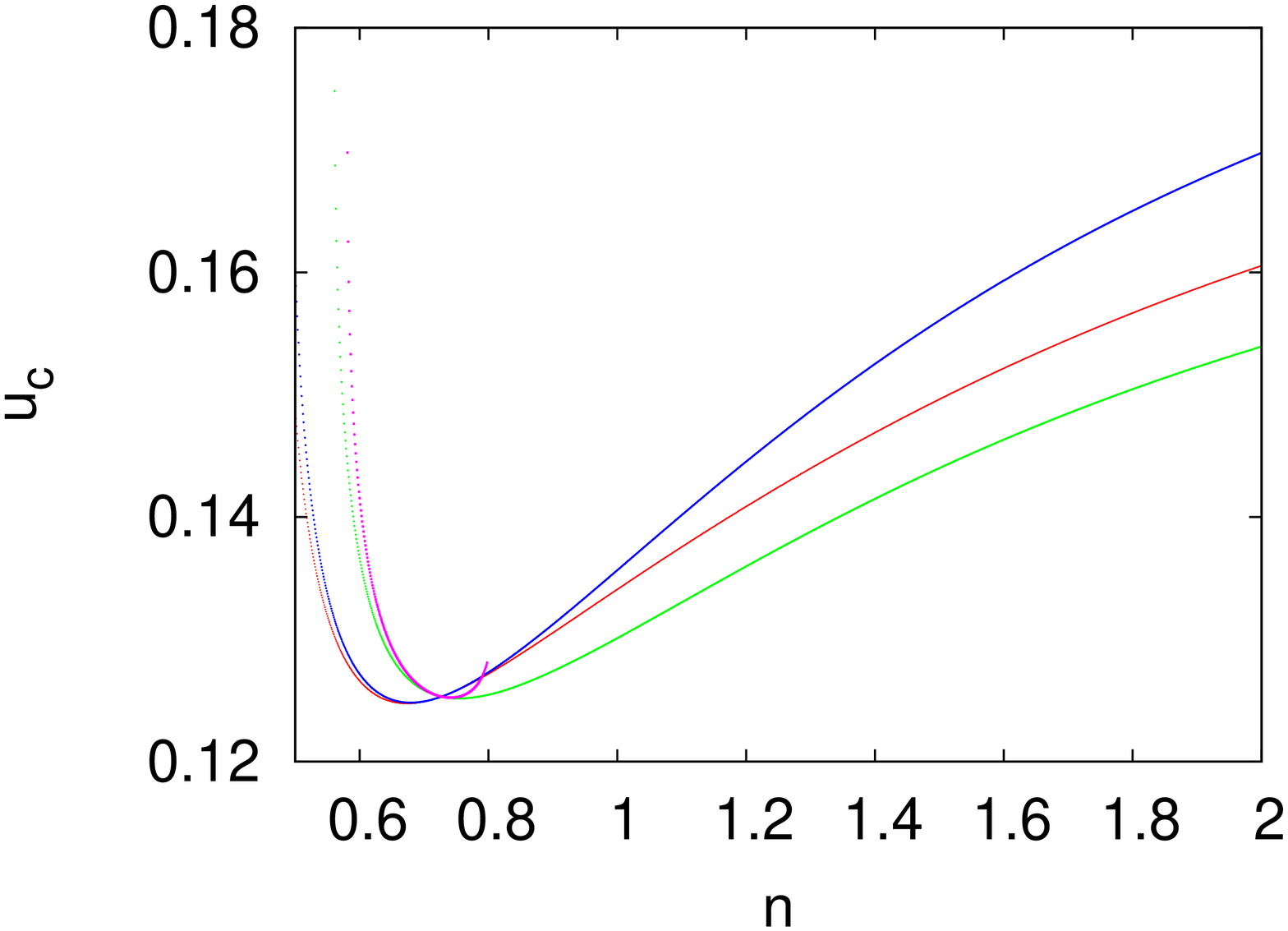}
}
\end{minipage}
\caption[]{Left: Schematic phase diagram in the variables $(u,h)$ of the series 
\eq(\ref{eq_u_h_mu}). Right: Poles of different Pad\'e approximants (red:[1,3], 
blue:[0,2], green:[0,3], purple:[1:2]) accumulate for a slope parameter $n=0.730$ 
at the critical point $u_c=0.126$.}\label{fig_schem}
\end{figure}

In order to locate the critical point $(\beta_c,\kappa_c)(\mu)$ we have to adjust our 
analysis methods to multiple variables. For a given number of flavours, the schematic
phase diagram is shown in \fig\ref{fig_schem}.
Our expansion is performed about $(u,h)=(0,0)$ and we now approach the critical
point along some straight line starting at the origin,
\begin{eqnarray}
u=n\cdot t,\nonumber\\
h=\frac{1}{n}\cdot t,
\label{eq_scaling}
\end{eqnarray}
whose slope is tuned by $n$ and t parametrises the distance from the critical point.
We then expect a scaling behaviour
\begin{eqnarray}
\chi_L(t)\sim\frac{1}{(t_c-t)^{\lambda}},
\end{eqnarray}
with some critical exponent $\lambda$. The critical point is in the 3d Ising universality 
class, but we do not know the scaling fields and variables. The Polyakov loop 
in general mixes contributions from both energy-like and magnetic field-like variables. When approaching the critical point
along a straight line from the origin, the larger exponent will dominate and in case of 3d Ising universality this is $\gamma$.

Our strategy now is to vary $n$ and calculate DLog-Pad\'es and singularities for each
value. 
If the axis misses the critical point, there is no scaling behaviour and we 
expect any real poles in $t$, and hence in $u$, of the Pad\'e approximants to be widely scattered.
As the critical point is approached, these poles
accumulate in a narrow window, as 
in our $SU(2)$ study, cf. \fig\ref{fig_schem}.
Thus, we estimate $t_c$ as the mean value over all Pad\'e 
singularities evaluated at that $n$ for which its standard deviation 
is minimal. As an error estimate we take the larger one of those two values, 
where the standard deviation is 1.5 times its minimum. With this method we find for $\mu=0$
\begin{eqnarray}
n=0.730(16),\qquad\overline{t_c}=0.172(4),\qquad
\overline{\lambda}=1.03(3).\label{eq_result}
\end{eqnarray}
Let us note that the standard deviations for both $t$ and $\lambda$ 
reach their minimum for the same slope parameter $n$. The critical exponent 
obtained with this method again underestimates $\gamma$ compared to the Ising 
exponent. Judging from our experience with $SU(2)$, we associate this with the 
truncation of the strong coupling series. 
From \eq(\ref{eq_scaling}) we obtain
\begin{eqnarray}
u_c=0.126(1),\qquad h_c=0.236(11).
\end{eqnarray}

To get more accurate results, we now employ biasing with the 3d Ising exponent 
$\gamma=1.237$, which leads to the improved values
\begin{eqnarray}
u_c=0.131(1)\quad\rightarrow\quad\beta_c=2.03(2)\qquad\qquad h_c=0.249(13).
\end{eqnarray}
Note that, to leading order in the hopping expansion, $\beta_c$ does not depend on $N_f$.
Inverting \eq(\ref{eq_hopping}) in case of $\mu=0$ we find
\begin{eqnarray}
\kappa=\frac{1}{2}\left(\frac{h}{2N_f}\right)^{\frac{1}{N_{\tau}}}.
\end{eqnarray}
This leads to the following results for $\kappa_c(N_f)$ on a $N_{\tau}=1$ lattice
\begin{eqnarray}
N_f=1:\qquad\kappa_c&=&0.062(4),\nn\\
N_f=2:\qquad\kappa_c&=&0.031(2),\nn\\
N_f=3:\qquad\kappa_c&=&0.021(1).
\end{eqnarray}
For these small values our leading order in the hopping expansion should be an excellent
approximation. This further justifies use of
the relation \cite{GK}
\begin{eqnarray}
\kappa=\frac{1}{2}\,e^{-ma},
\end{eqnarray}
which is valid for heavy quarks, to obtain the critical quark masses as
\begin{eqnarray}
N_f=1:\qquad m_c/T&=&2.08(7),\nn\\
N_f=2:\qquad m_c/T&=&2.78(7),\nn\\
N_f=3:\qquad m_c/T&=&3.17(10).
\label{masses}
\end{eqnarray}
The relative size is consistent with qualitative expectations. Since the presence
of finite mass quarks weakens the first order transition, more flavours of quarks should
have a stronger effect and thus a larger critical quark mass. 

\subsection{Critical point for $\mu\neq0$}

Next we turn on a chemical potential. \eq(\ref{eq_u_h_mu}) 
has already been obtained for finite chemical potential, 
all we have to do is to repeat the same steps to search for singularities 
as described in the previous analysis. 
Of particular interest is the movement of the critical masses $m_c$ with $\mu$.
\begin{figure}[ht]
\begin{center}
\includegraphics[width=0.6\textwidth]{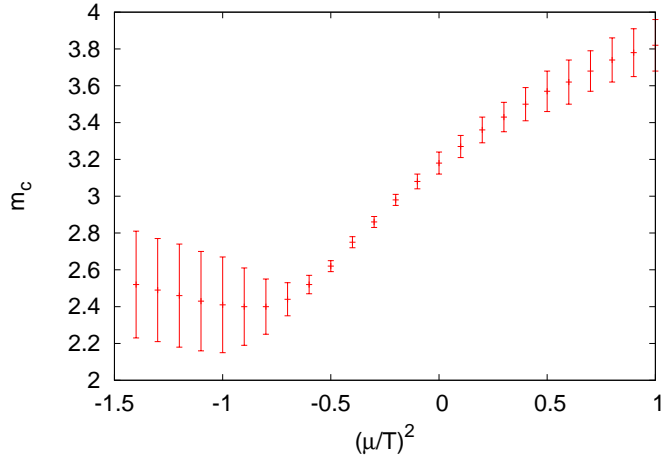}
\end{center}
\vspace*{-1.0cm}
\caption[]{The critical mass $m_c(\mu^2)$ for $N_f=3,N_\tau=1$ as a function of real and
imaginary chemical potential. Error bars are obtained with the same method as described before \eq(\ref{eq_result}).}
\label{fig_hc_mu}
\end{figure}
We have calculated this function for the case of $N_f=3$ at several points,
\fig \ref{fig_hc_mu}. 
The critical masses grow with real $\mu$, i.e.~the first order region is shrinking.
For small $\mu$ we performed a fit to a low order polynomial 
to get a rough picture of the behaviour of the signs of the different coefficients, 
\begin{eqnarray}
m_c(\mu^2)=3.18+0.94(1)\mu^2-0.34(1)\mu^4+0.037(18)\mu^6+\ldots,\label{eq_mc}
\end{eqnarray}
where the errors are those of the fitting procedure.
The shrinking of the first order region, and in particular the positive curvature of
the critical surface as well as the alternating signs
are in accord with the findings of a Monte Carlo investigation of the 
Potts model \cite{kfkt}. To leading order hopping expansion different $N_f$ only shift the constant term in \eq(\ref{eq_mc}).

The alternating signs indicate a convergence limiting singularity on the negative 
$\mu^2$ axis, i.e.~at imaginary chemical potential. 
Thus it is interesting to continue $m_c(\mu^2)$ also to negative values 
of the argument, by setting $\mu\rightarrow i\mu$. 
In \eq(\ref{eq_u_h_mu}), this means that $c$ now abbreviates $\cos(\mu)$ instead of 
$\cosh(\mu)$. The corresponding curve 
is also shown in \fig\ref{fig_hc_mu}. 
We observe a minimum at about $\mu^2\simeq-0.85$ and 
rapidly increasing errors with more negative $\mu^2$. 
We interpret this as the point where the singularity is located. 
This is fully consistent with the 
Roberge-Weiss $Z(N)$ transition point in the imaginary direction, which 
is known exactly to be at $\mu^2=-\pi^2/9\simeq -1.1$ \cite{roberge}.

\section{Conclusions}

We have explored the feasibility of calculating the critical couplings and exponents
of the deconfinement transition of lattice QCD by means of analytically computed 
strong coupling series, using the Polyakov loop susceptibility as an observable. 
Our treatment is purely analytical and goes beyond mean field theory.
Tables \ref{tab_crit}, \ref{tab_nt}, \eq(\ref{masses}) and \fig\ref{fig_hc_mu} 
contain our main results. 
For $SU(2)$ Yang-Mills theory on $N_{\tau}=1-4$ lattices, our results are
fully consistent with numerical values from Monte Carlo simulations.
In the cases $N_{\tau}=1,2$ we were able to reproduce these values nearly exactly with 
an accuracy of a few percent, for $N_{\tau}=3,4$ the results agree within 
increasing error bars. 
For $N_{\tau}>4$ the strong coupling series to the computed length becomes unpredictive, 
at least for our observable. A similar conclusion was drawn in \cite{Gr}.

In case of $SU(3)$ with heavy quarks we performed a strong coupling  
expansion in the effective action to leading order hopping expansion. 
Analysis of the $N_{\tau}=1$ series allowed to extract 
the critical quark mass $m_c(\mu)$, for which the first order deconfinement transition
goes critical before turning into a crossover. We find the first 
order transition region to shrink with increasing chemical potential. 
This is consistent with the findings from mean field theory \cite{Celik} and 
a Monte Carlo simulation of the effective theory with the same global symmetries,
the 3-state Potts model in 3 dimensions \cite{kfkt}. 
The latter is the appropriate 
effective model also for continuum QCD. 
Since our series furthermore correctly reflects the 
presence of the $Z(N)$ transition in
the direction of imaginary chemical potential, 
this would suggest that the qualitative phase
structure is correctly represented on lattices as coarse as $N_\tau=1$. 

We conclude that strong coupling expansions are able to provide qualitative and, on 
coarse lattices, quantitative information about the phase structure of lattice QCD
beyond mean field theory and the strong coupling limit, which easily 
extend to finite baryon density. 
While the location of phase transitions in the parameter space is subject to renormalisation
in the continuum limit, the presence of critical lines or surfaces is guaranteed 
by universality to survive in the continuum limit.
This strongly motivates further studies to 
extend our analyses to finer lattices and to the light quark regime.

\section*{Acknowledgement:}
We thank Ph.~de Forcrand and G.~M\"unster for numerous helpful discussions.
This work was supported by the BMBF project
{\em Hot Nuclear Matter from Heavy Ion Collisions and its Understanding from QCD}, 
No.~06MS254.


\begin{thebibliography}{99}

\bibitem{op}
O.~Philipsen,
  Eur.\ Phys.\ J.\ ST {\bf 152} (2007) 29
  [arXiv:0708.1293 [hep-lat]].

\bibitem{Kawamoto}
  N.~Kawamoto, K.~Miura, A.~Ohnishi and T.~Ohnuma,
  Phys.\ Rev.\  D {\bf 75}, 014502 (2007)
  [arXiv:hep-lat/0512023].

\bibitem{km}
  F.~Karsch and K.~H.~M\"utter,
  Nucl.\ Phys.\  B {\bf 313}, 541 (1989).


\bibitem{dff}
  M.~Fromm and P.~de Forcrand,
  arXiv:0811.1931 [hep-lat].

\bibitem{Wi}
  K.~G.~Wilson,
  Phys.\ Rev.\  D {\bf 10} (1974) 2445.

\bibitem{Mu}
  G.~M\"unster,
  Nucl.\ Phys.\  B {\bf 190}, 439 (1981)
  [Erratum-ibid.\  B {\bf 200}, 536 (1982),\ Erratum-ibid.\ B {\bf 205}, 648 (1982)].

\bibitem{drouffe}
  J.~M.~Drouffe and J.~B.~Zuber,
  Phys.\ Rept.\  {\bf 102}, 1 (1983).

\bibitem{PS}
  J.~Polonyi and K.~Szlachanyi,
  Phys.\ Lett.\  B {\bf 110}, 395 (1982).

\bibitem{GK}
  F.~Green and F.~Karsch,
  Nucl.\ Phys.\  B {\bf 238}, 297 (1984).

\bibitem{Po}
  A.~M.~Polyakov,
  Phys.\ Lett.\  B {\bf 72} (1978) 477.

\bibitem{Celik}
  T.~Celik, T.~Firat, Y.~Gunduc and M.~Onder,
  Phys.\ Rev.\  D {\bf 35}, 3958 (1987).

\bibitem{Faldt:1985ec}
  G.~F\"aldt and B.~Petersson,
  Nucl.\ Phys.\  B {\bf 265}, 197 (1986).


\bibitem{Su}
  L.~Susskind,
  Phys.\ Rev.\  D {\bf 20}, 2610 (1979).

\bibitem{Sv}
  B.~Svetitsky,
  Phys.\ Rept.\  {\bf 132}, 1 (1986).

\bibitem{Gr}
  F.~Green,
  Nucl.\ Phys.\  B {\bf 215}, 83 (1983).

\bibitem{BC}
  M.~Billo, M.~Caselle, A.~D'Adda and S.~Panzeri,
  Nucl.\ Phys.\  B {\bf 472}, 163 (1996)
  [arXiv:hep-lat/9601020].

\bibitem{Ohnishi}
  A.~Ohnishi, N.~Kawamoto and K.~Miura,
  J.\ Phys.\ G {\bf 34} (2007) S655
  [arXiv:hep-lat/0701024].

\bibitem{miura}
  A.~Ohnishi, K.~Miura, T.~Z.~Nakano and N.~Kawamoto,
  arXiv:0910.1896 [hep-lat].

\bibitem{nakano}
  K.~Miura, T.~Z.~Nakano, A.~Ohnishi and N.~Kawamoto,
  Phys.\ Rev.\  D {\bf 80}, 074034 (2009)
  [arXiv:0907.4245 [hep-lat]].


\bibitem{La}
  J.~Langelage, G.~M\"unster and O.~Philipsen,
  JHEP {\bf 0807}, 036 (2008)
  [arXiv:0805.1163 [hep-lat]].

\bibitem{BA}
  R.~Ben-Av, H.~G.~Evertz, M.~Marcu and S.~Solomon,
  Phys.\ Rev.\  D {\bf 44}, 2953 (1991).

\bibitem{Fi}
  J.~Fingberg, U.~M.~Heller and F.~Karsch,
  Nucl.\ Phys.\  B {\bf 392}, 493 (1993)
  [arXiv:hep-lat/9208012].

\bibitem{Ve}
  A.~Velytsky,
  Int.\ J.\ Mod.\ Phys.\  C {\bf 19}, 1079 (2008)
  [arXiv:0711.0748 [hep-lat]].

\bibitem{potts}
C.~Alexandrou, A.~Borici, A.~Feo, P.~de Forcrand, A.~Galli, F.~Jegerlehner and T.~Takaishi,
  Phys.\ Rev.\  D {\bf 60} (1999) 034504
  [arXiv:hep-lat/9811028].

\bibitem{ks}
F.~Karsch and S.~Stickan,
  Phys.\ Lett.\  B {\bf 488} (2000) 319
  [arXiv:hep-lat/0007019].

\bibitem{kfkt}
  S.~Kim, Ph.~de Forcrand, S.~Kratochvila and T.~Takaishi,
  PoS {\bf LAT2005}, 166 (2006)
  [arXiv:hep-lat/0510069].

\bibitem{mm}
  I.~Montvay and G.~M\"unster,
{\it  Cambridge, UK: Univ. Pr. (1994) 491 p. (Cambridge monographs on mathematical physics)}

\bibitem{gutt}
A.~J.~Guttman, in {\it Phase Transitions and Critical Phenomena}, Vol.~13, p.1, 
eds.~C.~Domb and J.~L.~Lebowitz, Academic Press, London, 1989

\bibitem{Ca}
  M.~Campostrini, A.~Pelissetto, P.~Rossi and E.~Vicari,
  Phys.\ Rev.\  E {\bf 65}, 066127 (2002)
  [arXiv:cond-mat/0201180].

\bibitem{hasenfratz}
  P.~Hasenfratz and F.~Karsch,
  Phys.\ Lett.\  B {\bf 125}, 308 (1983).

\bibitem{roberge}
  A.~Roberge and N.~Weiss,
  Nucl.\ Phys.\  B {\bf 275}, 734 (1986).



\end{thebibliography}
\end{document}